\documentclass[9pt,twocolumn,twoside]{osajnl}
\usepackage{siunitx}
\usepackage{amsmath}
\usepackage{amssymb,amsfonts}
\usepackage{multirow}
\usepackage{threeparttable}
\usepackage{cuted}
\usepackage{stfloats}

\journal{ol} % Choose journal (ao, aop, josaa, josab, ol, optica, pr)

\setboolean{shortarticle}{true}
\title{Reconfigurable Meta-Radiator Based on Flexible Mechanically Controlled Current Distribution in Three-dimensional Space}

\author[1,2]{Nan-Shu~Wu}
\author[1,2,6]{Su~Xu}
\author[1]{Xiao-Liang~Ge}
\author[1]{Jian-Bin~Liu}
\author[1]{Hang~Ren}
\author[3]{Kuiwen~Xu}
\author[4]{Zuojia~Wang}
\author[4]{Fei~Gao}
\author[1]{Qi-Dai~Chen}
\author[1,5,7]{Hong-Bo~Sun}

\affil[1]{State Key Laboratory of Integrated Optoelectronics, College of Electronic Science and Engineering, Jilin University, Changchun 130012, China}
\affil[2]{Zhejiang Provincial Key Laboratory of Advanced Microelectronic Intelligent Systems and Applications, Hangzhou 310027, China}
\affil[3]{Engineering Research Center of Smart Microsensors and Microsystems, Ministry of Education, School of Electronics and Information, Hangzhou Dianzi University, Hangzhou 310018, China.}
\affil[4]{Interdisciplinary Center for Quantum Information, State Key Laboratory of Modern Optical Instrumentation, Zhejiang University, Hangzhou 310027, China}
\affil[5]{State Key Laboratory of Precision Measurement Technology and Instruments, Department of Precision Instrument, Tsinghua University, Haidian, Beijing 100084, China}
\affil[6]{xusu@jlu.edu.cn}
\affil[7]{hbsun@tsinghua.edu.cn}

\begin{abstract}
In this paper, we provide an experimental proof-of-concept of this dynamic 3D current manipulation through a 3D-printed reconfigurable meta-radiator with periodically slotted current elements. By utilizing the working frequency and the mechanical configuration comprehensively, the radiation pattern can be switched among 12 states. Inspired by maximum likelihood method in digital communications, a robustness-analysis method is proposed to evaluate the potential error ratio between ideal cases and practice. Our work provides a previously unidentified model for next-generation information distribution and terahertz-infrared wireless communications.
\end{abstract}

\setboolean{displaycopyright}{true}

\begin{document}

\maketitle

% \section{Introduction}
Controlling electromagnetic waves and lights with a meta concept developed rapidly in the past decades. Initially, metamaterials, consisting of periodic three-dimensional (3D) meta-atoms with subwavelength scale, were studied to achieve constitutive parameters hardly found in natural materials \cite{peng2019transverse}. When the effective constitutive parameters of metamaterials become tensors, metamaterials could involve more degree of freedom (DOF) compared to bulk isotropic materials in nature. The higher DOF in constitutive relation will extend interaction between electromagnetic waves and the matters and bring more possibility to manipulate electromagnetic waves freely. With the development of metamaterials, various novel meta-devices have been proposed, such as invisibility cloaks\cite{schurig2006metamaterial,zheng2019experimental}, unconventional radiation control\cite{jing2019spiral,xu2017deep,li2020fib,chen2020189}, compact absorbers\cite{song2018transparent,dong2020solar,lv2021switchable}, sharp-corner benders\cite{zhou2019substrate,xu2015broadband}, and deep-subwavelength resolution lens\cite{sun2015experimental,sun2018toward}.

Different from bulk metamaterials, metasurfaces are another class of artificial electromagnetic structures characterized by the properties of surface current (e.g., the amplitude, phase shifting, and spatial distribution)\cite{sun2012gradient}. Compared to the homogeneous metallic/conducting surface, metasurfaces provide a two-dimensional current DOF for controlling the radiation of electromagnetic waves in compact dimension\cite{yu2014flat,ni2015ultrathin,qian2020deep,xu2020cross}. To achieve multifunctional integration in compact dimensions, the dynamic manipulation on electromagnetic radiation also attract notorious interest in the metasurface design. For instance, Prof. Cui and his colleagues come up with the concept of time-space digital coding metasurface, which successfully merges the boundary between digital and analogue systems and provides a new routine to next-generation wireless communications\cite{cui2014coding,ma2020information,nswu2020}.

From previous literatures, one can see that both of metamaterials and metasurfaces enrich their design DOFs by expanding the dimension of parameters. If we can merge the unique characteristics of metamateials and metasurfaces by involving the 3D DOF and 2D current manipulation together, an unidentified platform for metadevice design and multifunctional integration might be emerged.

In this work, we experimentally demonstrate the proof-of-concept of a 3D current distribution manipulation with the use of a flexible meta-radiator. The flexible meta-radiator is constructed by periodically arranged slots and fabricated through 3D-printing technique. Each slot in meta-radiator is equivalent to a current atom, whose spatial position, orientation and phase shift can be adjusted by switching the radiation frequency and the mechanical state of the meta-radiator. The 3D current distribution manipulation results a dynamic spatial beam distribution with 12 radiation states. To evaluate the dynamic radiation error that may potentially occurs in the practical application scenarios, a robustness-analysis method is proposed by comparing the Euclidean-like distance among various radiation states. Our work indicates manipulating current in a 3D space actively may introduce multi-dimensional distribution capacity in the photonic device design and provide an alternative platform for information distribution and communications.
% \section{Concept \& Structure}
\label{section2}
% 
% \subsection{Concept construction \& physical model}
%%==============Original 
% The concepts of the current manipulation in various ways are depicted in Fig. 1. For homogeneous conducting materials in nature (Fig.~\ref{fig1}), the in-plane current can be expressed by a simple vector, e.g., \(\vec{J}=\hat{x}J_{x}+\hat{y}J_{y}\), with uniform phase and amplitude under a given plane-wave excitation. In contrast to the case of homogeneous conducting surface, the in-plane vector current on a metasurface (Fig.~\ref{fig1}(b)) could be written by \(\vec{J}_{2 D}=\hat{x}J_{x}\left(x,y\right)+\hat{y}J_{y}\left(x,y\right)\), which indicates varying current forms a 2D in-plane DOF. Additionally, the current DOF can be extended from 2D to 3D by stacking single-layer metasurfaces into a multi-layer architecture, as shown in Fig.~\ref{fig1}(c). Under this architecture, the current distribution, with both of in-plane and out-plane components, should be amended to \(\vec{J}_{3 D}=\hat{x}J_{x}\left(x,y,z\right)+\hat{y}J_{y}\left(x,y,z\right)+\hat{z}J_{z}\left(x,y,z\right)\), which could involve more DOF compared to the cases in Figs.~\ref{fig1}(a) and Fig.~\ref{fig1}(b). Then, the remained question is how to manipulate this 3D current DOF in practice.
%%==============Original 
%%==============Modified
The concepts of the 3D current manipulation is depicted in Fig. 1. Within the depicted structure, the current distribution is with both in-plane and out-plane components, i.e., \(\vec{J}_{3 D}=\hat{x}J_{x}\left(x,y,z\right)+\hat{y}J_{y}\left(x,y,z\right)+\hat{z}J_{z}\left(x,y,z\right)\), which could involve more current distributions compared to the cases of 1D current manipulation (e.g., homogeneous metallic plate) and 2D one (e.g., metasurface). Then, the remained question is how to manipulate this 3D current distribution in practice.

\begin{figure}[htbp]
  \vspace{-1mm} 
  \centering
  \includegraphics[width=0.4\textwidth]{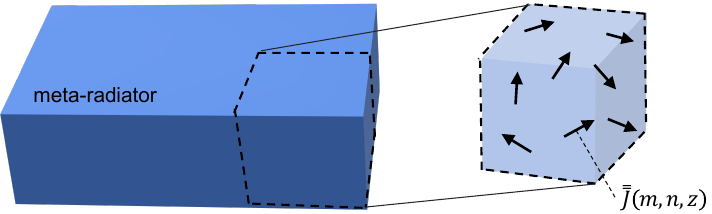}
  \caption{The 3D current manipulation in bulk meta-structure.}
  \label{fig1}
  \vspace{-2mm} 
\end{figure}

% \subsection{Physical realization for proof-of-concept}

\begin{figure}[htbp]
  \vspace{-1mm} 
  \centering
    \includegraphics[width=0.475\textwidth]{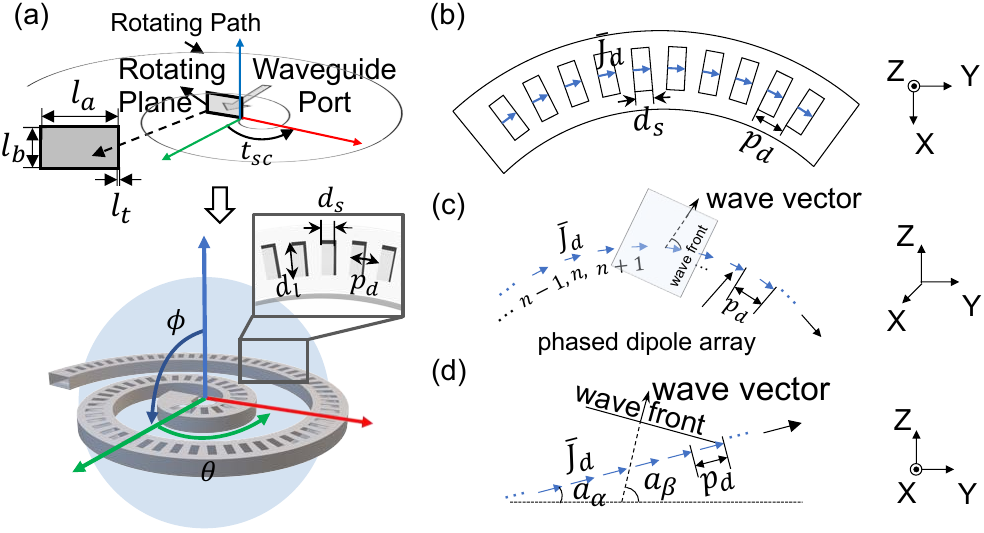}
  \caption{The schematic view of proposed meta-radiator. (a) The geometrical outline and the schematic view of spiral meta-radiator. (b) The equivalent method: each slot is seen as a current element with periodicity of \(p_d\). (c)-(d) The operating principle of 3-D current manipulating and reconfigurable radiation. Here \(\theta\) represents the polar angle and \(\phi\) is the azimuth angle, \(a_{\alpha}\) and \(a_{\beta}\) characterize the elevation angle and radiation angle to \(x-y\) plane.}
  \label{fig2}
  \vspace{-2mm} 
\end{figure}

Here, we draw lessons from slot antennas, which have been used as the current element for the Cherenkov radiation generation\cite{xi2009experimental}, to construct the 3D current manipulation. For the proof-of-concept, we arrange the current elements along a spiral curve
\begin{equation}
  L\left(t_{sc}\right):\left\{
    \begin{array}{lr}
      x_{sc}=d_{sc} t_{sc}\cos\left(t_{sc}\right), &  \\
      y_{sc}=d_{sc} t_{sc} \sin\left(t_{sc}\right), & {t_{sc}}_{min}\leq t_{sc}\leq {t_{sc}}_{max}\,,\\
      z_{sc}=e_{sc} t_{sc}, &  
    \end{array}
\right.
\end{equation}to stuff the space. As shown in Fig.~\ref{fig2}(a), \(d_{sc}\) is radial distance, \(e_{sc}\) contributes to the displacement in \(z\) direction, and \(t_{sc}\) is the parameter controlling the curve range. The tangent vector of \(L(t_{sc})\) at any given \(t_{sc}\) is \(\vec{L}^{\prime}\left(t_{sc}\right)=\hat{x}\frac{\partial x_{sc}}{\partial t_{sc}}+\hat{y}\frac{\partial y_{sc}}{\partial t_{sc}}+\hat{z}\frac{\partial z_{sc}}{\partial t_{sc}}=\left\{\frac{\partial x_{sc}}{\partial t_{sc}},\frac{\partial y_{sc}}{\partial t_{sc}},\frac{\partial z_{sc}}{\partial t_{sc}}\right\}\). The unit tangent vector of the 3D current path is \(\hat{s}=\frac{\vec{L}^{\prime}\left(t_{sc}\right)}{\left\lvert  \vec{L}^{\prime}\left(t_{sc}\right)\right\rvert }\), which is the direction vector in the Cartesian coordinate system. The periodic rectangular slots, with the size of \(d_l{\times}d_s\) and the period of \(p_d\) are arranged on the surface of spiral meta-radiator. Each slot can generate a displacement current \(\vec{J}\) along the tangential direction of the spirally curved rectangular waveguide, as shown in Fig.~\ref{fig2}(b). The current element along \(\hat{s}\) generated by a given slot can be given by\cite{kong2008electromagnetic,jiang2020backward}
\begin{equation}
  \vec{J}\left(\vec{r^\prime} \right)=\hat{s}Id_s\delta\left(\vec{r^\prime} \right)\,,
\end{equation}
where the \(\vec{r^\prime}\) marks the position of source, \(I\) is the amplitude of current, \(d_s\) is the width of the slot, and \(\delta\) is the Dirac delta function. The electric field generated by a given current element is 
\begin{equation}
  \begin{aligned}
    \vec{E}\left(\vec{r}\right)&=i\omega\mu\left[\overline{\overline{I}}+\frac{1}{k_s^2}\nabla\nabla\right]\cdot\iiint \mathrm{d}\vec{r^{\prime}}\frac{e^{ik_s\left\lvert \vec{r}-\vec{r^{\prime}}\right\rvert}}{4\pi \left\lvert \vec{r}-\vec{r^{\prime}}\right\rvert }\hat{s}Id_s\delta\left(\vec{r^{\prime}}\right)\,,
  \end{aligned}
\end{equation}where \(\omega\) is the angular frequency of the electromagnetic waves, \(\mu\) is the relative permeability, \(\overline{\overline{I}}\) is the unit dyad in dyadic Green's functions, and \(k_s\) is the wave number along \(\hat{s}\).

\begin{figure}[htbp]
  \vspace{-1mm} 
	\centering
	  \includegraphics[keepaspectratio,width=0.475\textwidth]{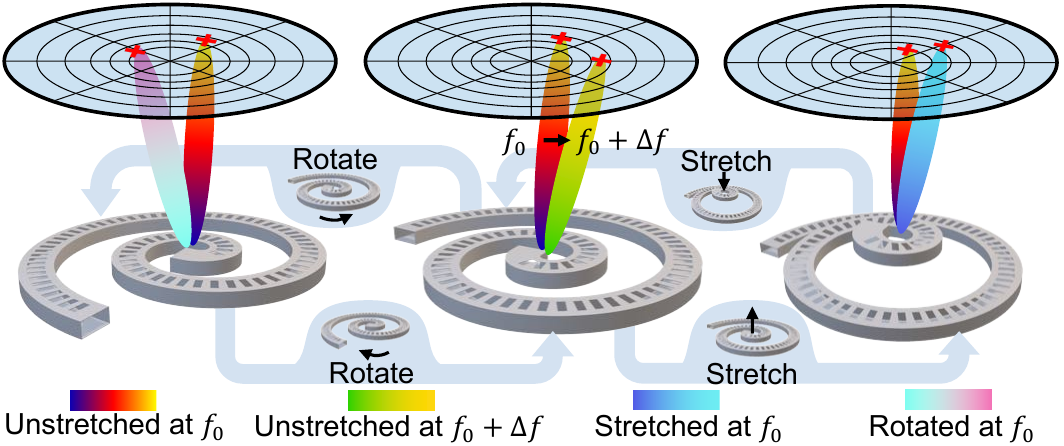}
	\caption{The schematic view of spatial-frequency-modulation-like meta-radiator. Each state can be switched by tuning working frequency (the middle part of the figure), implementing axial rotation on structure (the left and the middle part of the figure), and stretching vertically the structure (the middle and the right part of the figure).}
	\label{fig3}
  \vspace{-2mm} 
\end{figure}

For n-element current atoms distributed along the 3D spiral curve, the total current density can be formulated as
\begin{equation}
  \begin{split}
    \vec{J}_n\left(\vec{r^{\prime}}\right)&=\hat{s}Id_{s}\sum_{n = 1}^{\infty}e^{i n\alpha-\beta n d_{s}}\delta\left(x^{\prime}-d_{sc} n t_{sd}cos(n t_{sd})\right)\\
    &~~~~~~~~~~~~\delta \left(y^{\prime}-d_{sc}n t_{sd}sin(n t_{sd})\right)\delta \left(z^{\prime}-e_{sc}n t_{sd}\right)\,,\\
  \end{split}
\end{equation}where \(\alpha=k_sp_d\), \(\beta\) is the decay constant, and\(t_{sd}\) represents the radian element designed to decrease as \(n\) increases, leading to an unchanged \(p_d\) in the radius-varying curve. The relation between \(t_{sd}\) and \(p_d\) is \(\left(nd_{sc} t_{sd}^2\right)^2+\left( e_{sc} t_{sd}\right)^2=p_d^2\). Due to the spiral distributed current array, the total radiation ﬁeld can be written as
\begin{equation}
  \begin{aligned}
    &\vec{E}\left(\vec{r}\right)=i\omega\mu\left[\overline{\overline{I}}+\frac{1}{k_s^2}\nabla\nabla\right]\cdot\iiint \mathrm{d}\vec{r^{\prime}}\frac{e^{ik_{s}\left\lvert \vec{r}-\vec{r^{\prime}}\right\rvert}}{4\pi \left\lvert \vec{r}-\vec{r^{\prime}}\right\rvert }\hat{s}Id_{s}\\
    &~~~~~~~~~~\sum_{n = 1}^{\infty}e^{i n\alpha-\beta n d_{s}}\delta \left(x^{\prime}-d_{sc} n t_{sd}cos(n t_{sd})\right)\\
    &~~~~~~~~~~~~~~~~\delta \left(y^{\prime}-d_{sc}n t_{sd}sin(n t_{sd})\right)\delta\left(z^{\prime}-e_{sc} n t_{sd}\right)\,,\\
  \end{aligned}
\end{equation}

\noindent which is the superposition of the radiation fields of each current unit. From the previous equations, one can see that the total radiation highly depends on the properties of current elements. Since the spiral current distribution does not have axial symmetry and will introduce the eccentric radiation, the mechanical rotation will help to form the 2D current distribution. On the basis of this 2D current distribution, the out-of-plane current component could be easily realized by lifting the center of structure with a specific height. In short, curving and rotating the slot waveguide mechanically will enable an in-plane varying current component (Fig.~\ref{fig2}(c)) while stretching the meta-radiator vertically will enable the out-plane current distribution (Fig.~\ref{fig2}(d)). Additionally, switching excitation frequency can change the phase velocity of effective charge movement and phase shifting between adjacent current element, which results a frequency-dependent distribution. By combining the aforementioned mechanical operation and frequency switching together, the radiation state can be controlled in a multi-dimensional current distribution architecture. The projection of proposed reconfigurable radiation state can be mapped to constellation diagrams (Fig.~\ref{fig3}), just like the multi-dimensional modulation (e.g., frequency shift keying (FSK) modulation\cite{proakis2001digital}) in the area of telecommunications, for evaluating radiation characteristics. 

\begin{figure}[htbp]
    \centering
    \includegraphics[width=0.4\textwidth]{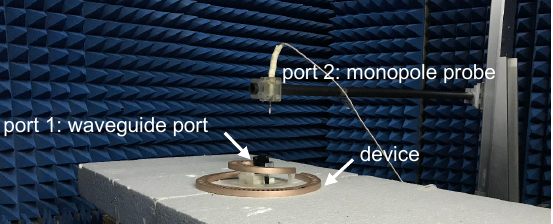}
\caption{Experimental setup.}
\label{fig4}
\end{figure}

% \section{Experimental validation and results}
\label{section3}
% \subsection{Experimental Setup}
In experiment, the meta-radiator is made by copper-coated PolyJet photopolymers via 3D printing technique. The working frequencies are set within the X band (\SIrange{8}{12}{\giga\hertz}). The parameters of the practical meta-radiator are cross-section length \(l_a=\)\SI{22.86}{\milli\metre}, cross-section width \(l_b=\)\SI{10.16}{\milli\metre}, structure thickness \(l_t=\)\SI{1}{\milli\metre}, \(d_l=\)\SI{10}{\milli\metre}, \(d_s=\)\SI{5}{\milli\metre}, \(p_d=\)\SI{10}{\milli\metre}, \(d_{sc}=\)\SI{10}{\milli\metre}, \(e_{sc}=0\), and \(t_{sc}\) is the rotating angle varying from \SIrange[]{-12}{0}{\radian}. The test environment is built up in a microwave anechoic chamber, as shown in Fig.~\ref{fig4}. A commercial vector network analyzer (VNA) Kesight E5071C is used in the measurement. Port 1 of the VNA is connected to the device via a waveguide adapter, and port 2 of the VNA is connected to the probe. A Linbou 3D automatic scanning platform NF03 is used to obtain the data over the \(x-y\) plane. The \(x\) component, \(y\) component, and \(z\) component of electric field are sampled to characterize the power distribution in the far-field region. A scanning area of \SI{600}{mm} × \SI{600}{mm} is defined with the step length of \SI{4}{\milli\metre} along \(x\) direction and \(y\) direction. The observation plane is set to \SI{190}{\milli\metre} above the structure.
\noindent
For simplicity, the data are captured from one rotating scenario (quadrant area) in both simulation and experiments, while data of scenarios can be achieved by manually adding a rotating angle to the coordinates. As the maximum value of the field amplitude is generated in the Cartesian coordinate system, a coordinate conversion is conducted from Cartesian coordinate to spherical for obtaining the 2D up-to-down vertical projection, and azimuth ticks are adjusted uniformly from \SIrange[]{0}{90}{\degree}. The radiation state distribution diagram is plotted after coordinate conversion and interpolation, for further performance analysis.
\begin{figure}[!t]
  \vspace{-1mm} 
  \centering
    \includegraphics[width=0.45\textwidth]{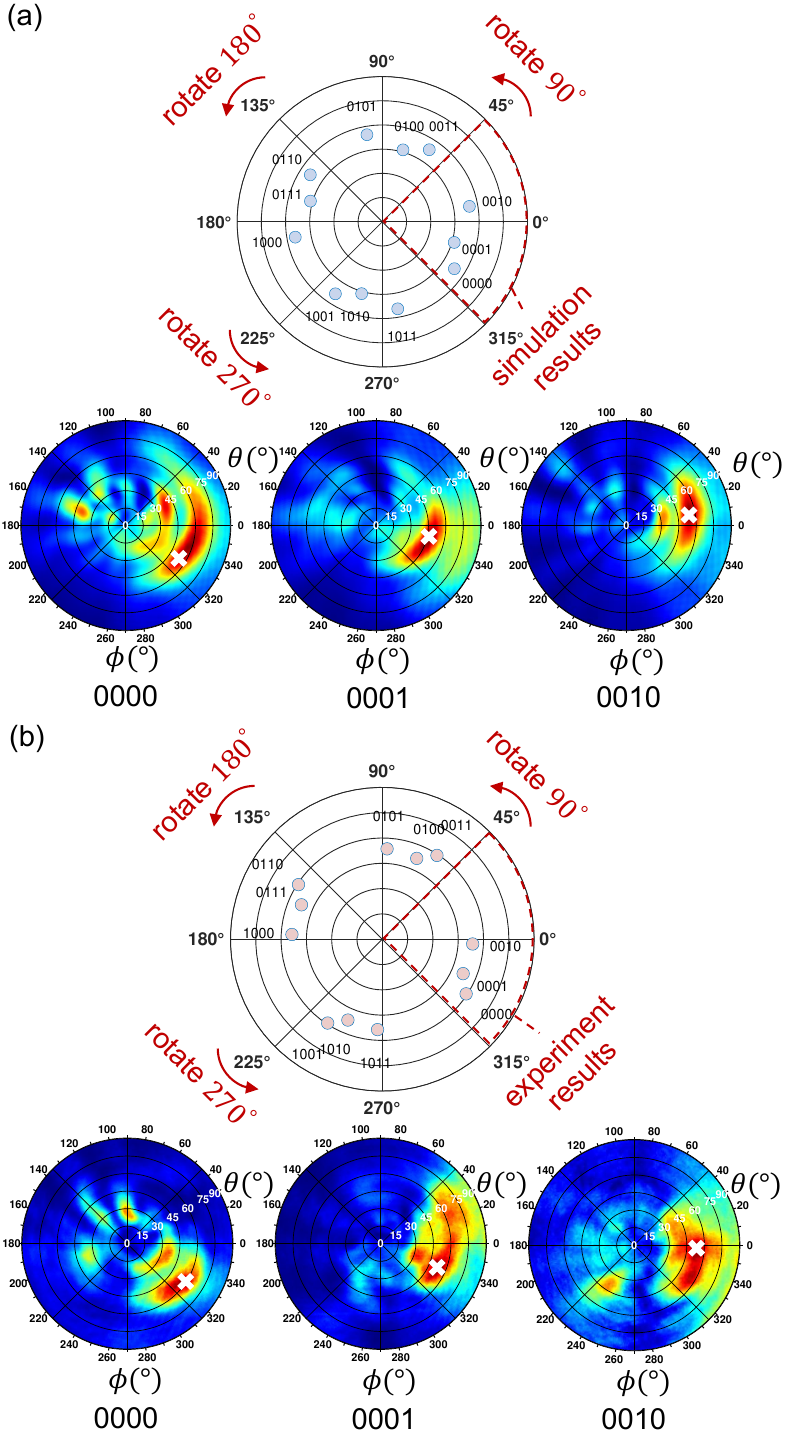}
  \caption{
    The constellation diagram and radiating field distribution. (a) Simulation; (b) Experiment. Blue/red dot represents the maximum electric field amplitude of separate scenarios for the case of simulation/experiments. Only the first three states are obtained through simulation and experiment, while other states are obtained by rotation directly.
  }
  \label{fig5}
  \vspace{-4mm} 
\end{figure}

In both simulation and experiment the constellation diagrams of the radiation states are shown in Fig.~\ref{fig5}. For the case of numerical simulation (Fig.~\ref{fig5}(a)), there are three radiation states in each defined quadrant area of the space. For example, States \(0000\), \(0001\), and \(0010\) exist in the quadrant area that ranges from \SIrange[]{315}{45}{\degree} counterclockwise. Here, State \(0000\) represents that radiation characteristic of frequency being \SI{11}{\giga\hertz} with un-stretched structure, while State \(0001\) is realized by stretching up structure manually with the height of \SI{40}{\milli\metre} at \SI{10}{\giga\hertz} in experiment. Similarly, tuning the frequency to \SI{10}{\giga\hertz} and keeping the structure without any stretching forms State \(0010\). Combining the mechanical rotation operation, the constellation diagram of radiation includes 12 states totally. The experimental results are presented as well, which is shown in Fig.~\ref{fig5}(b). For the simulation cases, the polar angles and azimuth angles (\(\phi, \theta\)) of States \(0000\), \(0001\), and \(0010\) are (330.90\si{\degree}, 54.18\si{\degree}), (343.96\si{\degree}, 46.35\si{\degree}) and (10.03\si{\degree}, 54.78\si{\degree}), respectively, while for the experiment, the corresponding spatial locations of aforementioned states are (326.90\si{\degree}, 53.28\si{\degree}), (336.94\si{\degree}, 52.07\si{\degree}) and (356.99\si{\degree}, 53.64\si{\degree}), respectively. For the selected states, the mismatch (\(\Delta\phi, \Delta\theta\)) between experiment and simulation are (4\si{\degree}, 0.9\si{\degree}), (7.02\si{\degree}, -5.72\si{\degree}) and (13.04\si{\degree}, 1.14\si{\degree}), respectively. Such mismatches could be caused by inaccurate manual operation and the structural distortion of soft-printed slot waveguide. Here, learning from the likelihood statistics, a Euclidean-distance-like factor is introduced to compare the mismatch significance between the practical cases and the ideal cases. That is,
\begin{equation}
  \begin{aligned}
    \label{eqdkm}
    d_{km}&=\left\lvert \hat{\psi}_{k}\left\langle\phi_k,\theta_k\right\rangle-\hat{\psi}_{m}\left\langle\phi_m,\theta_m\right\rangle\right\rvert \\
    &=\left[r_csin(\phi_k)^2-2r_c^2cos(\theta_k-\theta_m)sin(\phi_k)sin(\phi_m)\right.\\
    &~~~\left. +r_csin(\phi_m)^2\right]^{\frac{1}{2}}\,,
  \end{aligned} 
\end{equation} where \(\hat{\psi}_{k}\left\langle\phi_k,\theta_k\right\rangle\) and \(\hat{\psi}_{m}\left\langle\phi_m,\theta_m\right\rangle\), in a constellation diagram, represent spatial locations of radiation states. \(r_c\) is the radius of constellation diagram. For simplicity, \(r_c\) is set to \(1\). A shorter Euclidean distance indicates a greater likelihood between practical and ideal cases. Briefly, only the Euclidean distances of State \(0000\), \(0001\), and \(0010\) are selected for calculation as example, which are shown in table~\ref{tab2}. For example, the minima of Euclidean distances (\(0.057\)) for the experimental State \(0000\) occurs when it is calculated from its own simulation case. Similar to State \(0000\), the corresponding minimum Euclidean distances for State \(0001\) and State \(0010\), are \(0.113\) and \(0.185\), appearing when the likelihood is calculated compared to their own ideal radiation direction. Since the Euclidean-like distances between the simulation and experiment of each state are much shorter than the ones between mutual states, any state in the experiment can be identified to their own pre-defined state. Therefore, the mismatch between simulation and experimental results will not introduce a wrong recognition.

\newcommand{\tabincell}[2]{\begin{tabular}{@{}#1@{}}#2\end{tabular}}
\begin{table*}[tp]
  \vspace{-3mm} 
  \caption{Euclidean-like distance in simulation and in experiment}
  \centering
  \resizebox{.75\textwidth}{!}{ 
  \begin{tabular}{l l l l l l l l l l l l l l}
  % \multicolumn{14}{c}{\(\star\) horizontal array represents original and C4 symmetry induced experiment results, vertical array represents simulation results}  \\
  \toprule 
  \multirow{2}{*}{State\(\star\)}  & & \multirow{2}{*}{\(0000\)} & \multirow{2}{*}{\(0001\)} & \multirow{2}{*}{\(0010\)} & \multirow{2}{*}{\(0011\)} & \multirow{2}{*}{\(0100\)} & \multirow{2}{*}{\(0101\)} & \multirow{2}{*}{\(0110\)} & \multirow{2}{*}{\(0111\)} & \multirow{2}{*}{\(1000\)} & \multirow{2}{*}{\(1001\)} & \multirow{2}{*}{\(1010\)} & \multirow{2}{*}{\(1011\)}\\
  \\ \hline
  0000 & & \textbf{0.057} & 0.239 & 0.595 &  1.180  & 1.227  & 1.485  & 1.611  & 1.508 & 1.505 & 1.100 & 0.908 & 0.644\\  
  0001  & & 0.087 & \textbf{0.113}  & 0.458 &  1.070 & 1.134  & 1.412  & 1.597  & 1.509 & 1.539 & 1.189 & 1.003 & 0.766 \\ 
  0010  & & 0.365 & 0.192 & \textbf{0.185}&  0.856 & 0.954 & 1.270  & 1.574  & 1.519 & 1.612 & 1.371 & 1.198 & 1.010\\
  \bottomrule 
  \multicolumn{14}{c}{\(\star\) \tabincell{c}{Vertical array represents simulation results, and the horizontal array represents the experiment results.}} \\
  \end{tabular}
  }
  \label{tab2}
  \vspace{-3mm} 
\end{table*}

This likelihood analysis will provide great prospects for future terahertz-infrared wireless communications for specialized scenarios and harsh conditions. For instance, the additional data that include the information of radiation states can be packaged into the coding flow. Such additional data that includes transmitter's properties can be used as the encryption key of the coding flow and could be decoded at the receiving end through the aforementioned Euclidean distance estimation. Without the pre-defined encoding encryption information, the transmitted information will be difficult to be cracked. By using this encryption architecture, the wireless communications security will be enhanced at the hardware level, in addition to the software level.
% On the other hand, this Euclidean distance method will also help to track the position disturbance emission source through changes in the retrieved constellation diagram, which may find application in remote sensing. For example, the movement of the Antarctic ice sheet or the occurrence of land subsidence could be discovered and analyzed if the states received by monitoring satellite vary due to changes in the 3D current DOF. It could be expected as an economical and efficient method of detecting global climate change.
% \section{Conclusion \& Discussion}
% \label{section4}

In conclusion, we experimentally demonstrate a reconfigurable spiral meta-radiator based on the concept of 3D current manipulation. A constellation diagram with 12 radiation states was achieved and a robustness method was introduced to evaluate the likelihood and correlation between different radiation states. Compared to the previous reconfigurable radiating devices, our work not only presents the advantages of high radiation tunability in metasurfaces\cite{cui2014coding,li2017controlling,zhao2020recent}, but also provides the proof-of-concept of 3D current distribution manipulation as the complementary solution of present meta-devices. 
% Our work meets the present technical trend of security communication in practical scenarios. For example, in satellite communications, the antenna's position and orientation can be used as part of the key. When the varying communication encoding key is pre-identified with the form of reconfigurable radiation states, the hackers cannot easily obtain either the temporal or spatial information of high-DOF radiation distribution at the same time. In this way, it is very difficult for hackers to retrieve the correct key and information through the single-node monitoring, and consequently the probability of decryption will be decreased significantly. Therefore, our work may find potential applications in 6G satellite-network based communication and multi-channel key distribution.
\begin{backmatter}
\bmsection{Funding} National Natural Science Foundation of China (61805097, 61935015, 61825502, 61971174, 61801426, 61801268); National Key R\&D Program of China (2017YFB1104300); ZJNSF (Z20F010018).
\vspace{-1mm} 
\bmsection{Disclosures} The authors declare no conflicts of interest.
\vspace{-1mm} 
\bmsection{Data Availability} Data underlying the results presented in this paper are not publicly available at this time but may be obtained from the authors upon reasonable request.
\end{backmatter}

\vspace{-3mm} 

% Bibliography
\bibliography{references}

% Full bibliography added automatically for Optics Letters submissions; the following line will simply be ignored if submitting to other journals.
% Note that this extra page will not count against page length
\bibliographyfullrefs{references}

%Manual citation list
%\begin{thebibliography}{1}
%\bibitem{Zhang:14}
%Y.~Zhang, S.~Qiao, L.~Sun, Q.~W. Shi, W.~Huang, %L.~Li, and Z.~Yang,
 % \enquote{Photoinduced active terahertz metamaterials with nanostructured
  %vanadium dioxide film deposited by sol-gel method,} Opt. Express \textbf{22},
  %11070--11078 (2014).
%\end{thebibliography}

% Please include bios and photos of all authors for aop articles
\ifthenelse{\equal{\journalref}{aop}}{%
\section*{Author Biographies}
\begingroup
\setlength\intextsep{0pt}
\begin{minipage}[t][6.3cm][t]{1.0\textwidth} % Adjust height [6.3cm] as required for separation of bio photos.
  \begin{wrapfigure}{L}{0.25\textwidth}
    \includegraphics[width=0.25\textwidth]{john_smith.eps}
  \end{wrapfigure}
  \noindent
  {\bfseries John Smith} received his BSc (Mathematics) in 2000 from The University of Maryland. His research interests include lasers and optics.
\end{minipage}
\begin{minipage}{1.0\textwidth}
  \begin{wrapfigure}{L}{0.25\textwidth}
    \includegraphics[width=0.25\textwidth]{alice_smith.eps}
  \end{wrapfigure}
  \noindent
  {\bfseries Alice Smith} also received her BSc (Mathematics) in 2000 from The University of Maryland. Her research interests also include lasers and optics.
\end{minipage}
\endgroup
}{}

\end{document}